\documentclass[letterpaper,11pt]{article}
\usepackage{amssymb}
\usepackage{amsmath}
\usepackage{amstext}
\usepackage{graphicx,epsfig}
\usepackage{epsfig}
\usepackage{verbatim} 
\usepackage{color}
\usepackage{ulem}
\usepackage{enumitem}
\usepackage{subfigure}
\usepackage{bbm}
\usepackage{parskip}
\usepackage{dsfont}
\usepackage[numbers,sort&compress]{natbib}
\usepackage[body={6.5in, 9in}, right=1in, top=1in]{geometry}
\usepackage{mathtools}
\usepackage{comment}
\usepackage{caption}
\usepackage{bbold}
\usepackage{float}
\usepackage{tabularx}
\usepackage{makecell}

\newcommand{\Comment}[1]{{}}
\definecolor{darkblue}{rgb}{0.15,0.35,0.55}
\definecolor{reddish}{rgb}{0.65, 0.2, 0.2}
\usepackage[linktocpage=true]{hyperref}
\hypersetup{
colorlinks=true,
citecolor=darkblue,
linkcolor=reddish,
urlcolor=darkblue,
pdfauthor={},
pdftitle={},
pdfsubject={}
}

\newcommand{\be}{\begin{equation}}
\newcommand{\ee}{\end{equation}}
\newcommand{\bea}{\begin{eqnarray}}
\newcommand{\eea}{\end{eqnarray}}
\newcommand{\beas}{\begin{eqnarray*}}
\newcommand{\eeas}{\end{eqnarray*}}

\def\({\left(}
\def\){\right)}

\def\gsim{ \lower .75ex \hbox{$\sim$} \llap{\raise .27ex \hbox{$>$}} }
\def\lsim{ \lower .75ex \hbox{$\sim$} \llap{\raise .27ex \hbox{$<$}} }

\usepackage{xcolor,colortbl}

\begin{document}
\def\thefootnote{\fnsymbol{footnote}}

\begin{center}
\LARGE{\textbf{Baryon-Interacting Dark Matter: heating dark matter and the emergence of galaxy scaling relations}} \\[0.5cm]
 
\large{Benoit Famaey,${}^{\rm a,}$\footnote{\href{mailto:benoit.famaey@astro.unistra.fr}{\texttt{benoit.famaey@astro.unistra.fr}}} Justin Khoury,${}^{\rm b,}$\footnote{\href{mailto:jkhoury@sas.upenn.edu}{\texttt{jkhoury@sas.upenn.edu}}} Riccardo Penco,${}^{\rm c,}$\footnote{\href{mailto:rpenco@andrew.cmu.edu}{\texttt{rpenco@andrew.cmu.edu}}} and Anushrut Sharma${}^{\rm b,}$\footnote{\href{mailto:anushrut@sas.upenn.edu}{\texttt{anushrut@sas.upenn.edu}}}}
\\[0.5cm]

\small{
\textit{$^{\rm a}$ Universit\'e de Strasbourg, CNRS, Observatoire astronomique de Strasbourg,\\ UMR 7550, F-67000 Strasbourg, France}}

\vspace{.2cm}

\small{
\textit{$^{\rm b}$ Center for Particle Cosmology, Department of Physics and Astronomy, \\ University of Pennsylvania, Philadelphia, PA 19104}}

\vspace{.2cm}

\small{
\textit{$^{\rm c}$ McWilliams Center for Cosmology, Department of Physics, \\ Carnegie Mellon University, Pittsburgh, PA 15213}}

\vspace{.2cm}

\end{center}

\vspace{.6cm}

\hrule \vspace{0.2cm}
\centerline{\small{\bf Abstract}}
\vspace{-0.2cm}
{\small\noindent The empirical scaling relations observed in disk galaxies remain challenging for models of galaxy formation. The most striking among these is the Mass Discrepancy-Acceleration Relation (MDAR), which encodes both a tight baryonic Tully-Fisher relation (BTFR) and the observed diversity of galaxy rotation curves through the central surface density relation (CSDR). Building on our earlier work~\cite{Famaey:2017xou}, we propose here that the MDAR is the result of interactions between baryons and `Baryon-Interacting Dark Matter' (BIDM), which heat up the dark matter. Following a bottom-up, hydrodynamical approach, we find that the MDAR follows if: $i)$~the BIDM equation of state approximates that of an ideal gas; $ii)$~the BIDM relaxation time is order the Jeans time; $iii)$~the heating rate is inversely proportional to the BIDM density. Remarkably, under these assumptions the set of hydrodynamical equations together with Poisson's equation enjoy an anisotropic scaling symmetry. In the BIDM-dominated regime, this gives rise to an enhanced symmetry which fully captures the low-acceleration limit of the MDAR. We then show that, assuming a cored pseudo-isothermal profile at equilibrium, this set of equations gives rise to parameters reproducing the MDAR. Specifically, in the flat part of the rotation curve the asymptotic rotational velocity matches the parametric dependence of the BTFR. Moreover, in the central region of high-surface brightness galaxies, the profile reproduces the CSDR. Finally, by studying the time-dependent approach to equilibrium, we derive a global combination of the BTFR and CSDR, which matches the expectations in low surface-brightness galaxies. The form of the heating rate also makes model-independent predictions for various cosmological observables. We argue that our scenario satisfies existing observational constraints, and, intriguingly, offers a possible explanation to the EDGES anomaly. 
%We discuss promising avenues and potential challenges for embedding our scenario in a concrete particle physics model.
} 
\vspace{0.3cm}
\noindent
\hrule
\def\thefootnote{\arabic{footnote}}
\setcounter{footnote}{0}

\section{Introduction}

The nature of the dark sector of the Universe is certainly one of the most important questions of modern physics. Over the years, a picture has emerged in which the Universe is composed of~$\sim 5$\% baryonic matter,~$\sim 25$\% cold dark matter (CDM)---which for all practical purposes does not interact with itself or with baryons---and the rest by a cosmological constant~$\Lambda$. While this~$\Lambda$CDM model is very successful on large scales, a few tensions remain. 

On cosmological scales, one notable tension is the value of the Hubble constant as inferred from the Cosmic Microwave Background (CMB)---which has drifted towards smaller values together with a larger matter density~$\Omega_{\rm m}$ with better successive data from the WMAP mission~\cite{Hinshaw:2012aka}, and even more so after Planck~\cite{Aghanim:2018eyx}---to be contrasted with the higher value obtained from measurements of Type~Ia supernovae and lensing time-delays~\cite{Riess:2018byc,Birrer:2018vtm}. Whether this tension might be resolved through understanding systematics or whether it is a sign of new physics is still under debate (see~\cite{Knox:2019rjx} and references therein). Meanwhile, an interesting anomaly has surfaced around redshift~$z \sim 20$, where the EDGES experiment has reported an anomalously strong absorption in the measured 21~cm signal~\cite{Bowman:2018yin}. If not due to foreground contamination, this signal might indicate an over-cooling of the HI gas with respect to standard expectations, or a modification of the soft photon background beyond the CMB contribution.

On galactic scales, a number of observational challenges to the standard~$\Lambda$CDM model have also been actively debated in recent years, as galactic observations and numerical simulations of galaxies have improved in tandem~\cite{Bullock:2017}. Galaxy formation and evolution are processes that happen on~$\sim {\rm kpc}$ scales, where the physics of baryons can play a major role through gravitational feedback in modifying the quasi-equilibrium configuration of CDM on secular timescales. 

The most interesting challenge is that baryons and dark matter (DM) in galaxies seem to conspire in ways that were {\it a priori} unexpected, giving rise to tight scaling relations. The most famous such scaling relation is the baryonic Tully-Fisher relation (BTFR)~\cite{McGaugh:2000sr,Papastergis:2016jqv,LelliBTFR,Lelli:2019igz}, relating the fourth power of the asymptotic circular velocity of disk galaxies to their baryonic mass,~$V_{\rm flat}^4 \sim M_{\rm b}$. Interestingly, when matching the mass-function of DM halos to the luminosity function of galaxies---a procedure known as abundance matching (AM)---one gets a stellar-to-halo mass relation that nicely reproduces the normalization of the BTFR~\cite{DiCintio:2015eeq,Desmond:2017}, especially at baryonic masses around~$10^{10}\,{\rm M}_\odot$. 

However, as shown by~\cite{Desmond:2017} using Navarro-Frenk-White (NFW) profiles~\cite{NFW} for the assigned DM halos, the AM-predicted curvature of the BTFR is at odds with the data. This might be attributable to large uncertainties in AM at low masses, but is definitely problematic at high masses (above stellar masses of $\sim 10^{11} M_\odot$) where AM systematically overpredicts the halo mass of disk galaxies~\cite{Posti:2019ovz}. Furthermore, the observed small intrinsic scatter (only~$\sim 0.025$~dex for the orthogonal scatter) of the BTFR is in 3.6$\sigma$ disagreement with AM expectations~\cite{Desmond:2017}. While some outliers to the BTFR at the low and high-mass ends have been recently pointed out~\cite{Pina:2019rer,Ogle_2019} and still need to be confirmed by more observations due to possible systematics ({\it e.g.}, on the inclination at the low mass end) or unknowns ({\it e.g.} on the asymptotic flat velocity and on the total gas mass at the high mass end), the tightness of the BTFR for the bulk of low-$z$ high-quality galaxy rotation curves remains challenging in the~$\Lambda$CDM context.

Another aspect of the baryon-DM conspiracy is the {\it diversity} of rotation curves. Galaxies with the same asymptotic circular velocity---hence ``twins" of identical total baryonic mass on the BTFR---can display a broad range of rotation curve {\it shapes}, consistent with central DM densities ranging from cuspy NFW-like central profiles as predicted in DM-only simulations, to very large, constant-density cores of DM~\cite{Oman:2015xda}. There is in fact a positive correlation between the average DM density within 2~kpc and the baryon-induced rotational velocity at that radius~\cite{Ghari:2018jkc}. The circular velocity slope close to the center is thus directly correlated to the surface density of baryons. In other words, the rotation curve shapes of late-type spiral galaxies are all similar when expressed in units of disk scale-length~\cite{Swaters:2009by}, and the DM core size correlates with scale-length~\cite{Donato:2004}. 

Another way to express this correlation is the {\it central surface density relation}~\citep[CSDR,][]{Lelli:2016uea} between the central surface density of stars and the central dynamical surface density, related to the slope of the rotation curve. For small disk galaxies dominated by DM, the expectation {\it a priori} would have been instead that galaxies at a given maximum velocity scale display {\it similar} rotation curves because they should be embedded in similar DM halos. Thus this can be considered as a strong version of the old ``core-cusp" problem~\cite{deBlok:2009sp}. 

The diversity of galaxy rotation curves at a given velocity scale, their uniformity at a given baryonic surface density scale, together with the BTFR, can be summarized through what is nowadays known in disk galaxies as the {\it Radial Acceleration Relation} (RAR), or more generally as the {\it Mass Discrepancy-Acceleration Relation} (MDAR). This encodes a unique observational relation between the total gravitational field and the Newtonian acceleration generated by baryons at every radius~\cite{Sanders1990,McGaugh:2004aw,Gentile:2011,McGaugh:2016leg,Lelli:2017vgz}. 

While the general shape of the MDAR might be a natural outcome of~$\Lambda$CDM~\cite{DiCintio:2015eeq,Keller:2017,Navarro:2017,Ludlow:2017,Read:2016}, and despite debates on its universality~\cite{Rodrigues2018,McGaugh2018,Kroupa2018}, its normalization and very small scatter, the latter which could be entirely accounted for by observational errors on the inclination and distance of galaxies, remain puzzling~\cite{Li:2018tdo}. For instance, it has recently been argued that feedback becomes efficient at a characteristic acceleration scale similar to the one present in the MDAR, thereby explaining the transition from baryon-dominated to DM-dominated regimes in the MDAR~\cite{Grudic_2019stellar}. While interesting, this does not {\it per se} explain the details of the diversity of rotation curves encoded in the tightness of the MDAR, which should be related to the subtleties of the core-cusp transformation process. 

While the MDAR reduces to the BTFR in the flat part of rotation curves, the fact that galaxies obey the BTFR does not {\it a priori} imply that they will obey the MDAR in the rising parts of rotation curves. The fact that they do observationally is at the root of the diversity problem, as shown in~\cite{Ghari:2018jkc}. As reported by~\cite{Oman:2015xda}, feedback in cosmological hydrodynamical simulations from the EAGLE and APOSTLE projects is unable to produce large constant-density cores of DM as required by the data in a significant fraction of low-mass disk galaxies. On the other hand, the recent NIHAO simulations~\cite{Santos_2017,Dutton:2019gor} are much more efficient at forming cores and predict a tight MDAR, but in turn have problems at reproducing the most cuspy, steeply rising rotation curves~\cite{Ghari:2018jkc}. This illustrates that the effect of feedback on the central DM distribution in various cosmological hydrodynamical simulations is still far from settled, and that reproducing in detail the observed diversity of rotation curve shapes {\it together} with a tight MDAR still raises an interesting challenge for simulations of galaxy formation.

\subsection{Approaches to the MDAR}

In this context, it is natural to explore whether the above challenges find their root in a modification of the fundamental nature of DM. Alternatives to~$\Lambda$CDM exploring different properties of the DM sector are usually concerned with changing the DM particle mass~\cite{Bode:2001} or self-interactions~\cite{Spergel:1999mh}. Interactions with photons~\cite{Schewtschenko:2015} or neutrinos~\cite{Boehm:2017} in the early Universe have also been considered, affecting the linear regime. In self-interacting DM some recent encouraging results have shown how underdense halos can indeed be associated with extended baryonic disks~\cite{Kamada:2016,Creasey:2017,Ren:2018jpt}, in line with the trend of the MDAR. While very encouraging, rotation curve fits are still made with two parameters~\cite{Ren:2018jpt} and do not fully explain {\it ab initio} the tightness of the BTFR, as well as, {\it e.g.}, its tension with AM at high masses.

Given the tight correlation between the Newtonian gravitational field generated by baryons and the total gravitational field, the most direct and also most radical alternative explanation is that the gravitational law is, at least effectively, modified in galaxies~\cite{Bekenstein:1984tv,Bekenstein:2004ne,Skordis:2019fxt}. This paradigm, known as Modified Newtonian Dynamics (MOND)~\cite{Sanders:2002pf,Famaey:2011kh} was proposed almost ~40 years ago by Milgrom~\cite{Milgrom:1983ca,Milgrom:1983pn}. Within this framework, the MDAR was actually {\it predicted} well before it was precisely assessed by observations. The challenge with this approach is to reproduce the large-scale successes of~$\Lambda$CDM, in particular the exquisitely-measured CMB temperature anisotropies. There are additional challenges with the mass discrepancy in galaxy clusters~\cite{Angus:2007,Famaey:2011kh}, subgalactic scales~\citep[{\it e.g.},][]{Ibata:2011ri}, as well as solar system constraints~\cite{Hees:2014kta,Hees:2015bna} (though see~\cite{Babichev:2011kq}).

Less radical is the idea that DM acts as CDM cosmologically but generates an effective modification of gravity on galactic scales through long-range interactions~\cite{Blanchet:2006yt,Blanchet:2008fj,Khoury:2014tka,Blanchet:2015sra,Blanchet:2015bia}. A recent prototypical example in this category is based on DM superfluidity~\cite{Berezhiani:2015pia,Berezhiani:2015bqa,Khoury:2016ehj,Addazi:2018ivg,Khoury:2016egg,Fan:2016rda,Hodson:2016rck,Cai:2017buj,Berezhiani:2017tth,Hossenfelder:2018iym,Sharma:2018ydn,Alexander:2018fjp,Ferreira:2018wup,Berezhiani:2018oxf,Berezhiani:2019pzd}. More radical approaches include Modified DM~\cite{Ho:2010ca,Ho:2011xc,Ho:2012ar,Edmonds:2017zhg} and Verlinde's emergent gravity~\cite{Verlinde:2016toy,Hees:2017}, both inspired by gravitational thermodynamics. All such approaches boil down to some version of MOND on galaxy scales. 

Another route, as yet very much unexplored, is that the tight conspiracy between the distribution of baryons and the gravitational field in galaxies is the outcome of relatively short-range interactions between baryons and DM, which reorganize the DM distribution in the desired way without effectively modifying gravity.

In~\cite{Famaey:2017xou} we proposed a novel mechanism along these lines. The idea put forward was that the desired DM profile may naturally emerge as the equilibrium configuration resulting from DM-baryon short-range (collisional) interactions. This required replacing the traditional collisionless Boltzmann equation describing the DM fluid by a collisional Boltzmann transport equation with two fluids. The first and second order moments of this equation yield respectively the traditional Jeans' equation (akin to hydrostatic equilibrium) and a heat transport equation describing the exchange of energy between baryons and DM.  For static and isotropic configurations, the heat equation implies an actual equilibrium between the divergence of the heat flux within the DM fluid and the heating rate due to baryons. By retro-engineering the observationally-inferred knowledge of the MDAR in rotationally-supported disk galaxies, it was shown that an equilibrium configuration reproducing the MDAR can be attained if: i)~the heating rate is inversely proportional to the DM density; and ii)~if the relaxation time of DM particles is comparable to the dynamical time. 

Specifically, in~\cite{Famaey:2017xou} we concentrated on collisional interactions between heavy DM particles and baryons, in which baryons effectively cooled the DM medium. We could then demonstrate that, as long as the BTFR was obeyed at large radii, the MDAR would be satisfied at all radii. While setting the stage for follow-up studies, our original model suffered from a few important caveats. Firstly, the BTFR had to be assumed at equilibrium, and it was unclear how it might be achieved in the time-dependent case. Secondly, since the mechanism relied on cooling the DM fluid to reach equilibrium, one would need to start from relatively hot initial conditions, in contradiction with the successes of~$\Lambda$CDM on large scales, or, alternatively, the center of DM halos would need to be strongly up-scattered by very efficient feedback before being allowed to cool again. An additional concern is that the cooling mechanism could lead, in self-consistent simulations, to flattened DM halos or prominent dark disks, once halos have an initial spin. Finally, we assumed that we could coarse-grain the baryonic and DM distribution functions over a typical scale of a few pc, which cannot be the case for purely collisional interactions between DM particles and stars without strongly enhancing the DM density around stars.

\subsection{Baryon-interacting DM} 

In this paper we build on and further develop the original scenario of~\cite{Famaey:2017xou} in several crucial ways. Most importantly, instead of baryon-DM interactions cooling the DM medium,
we now focus exclusively on the case where the DM fluid is {\it heated} by baryons. This is {\it a priori} more desirable from the point of view of galaxy formation, since DM heating can
transform cusps into cores in central regions of galaxy halos. It also avoids the concern of forming flattened halos or dark disks. A second key difference pertains to the form of DM-baryon interactions. Whereas our original analysis~\cite{Famaey:2017xou} focused exclusively on short-range particle-particle collisions between DM and baryons, in the present analysis we remain general about the form of such interactions, which could happen on a pc-range. 

The basic framework is otherwise similar to~\cite{Famaey:2017xou}. After reviewing the MDAR in Sec.~\ref{MDAR review}, we set up in Sec.~\ref{setup} a
bottom-up approach to identify phenomenologically the kind of DM-baryon interactions necessary to reproduce the MDAR. By taking the first few velocity
moments of a collisional Boltzmann transport equation, we obtain a hydrodynamical description of DM governed by a continuity equation, a Jeans' or momentum equation, and,
crucially, a heat equation describing energy exchange between DM and baryon components. These are supplemented by the standard Poisson equation determining
the gravitational field.

The microphysics of DM is encoded in three physical quantities. The first quantity is the DM equation of state,~$P = P(\rho,v)$, specifying
the pressure as a function of density~$\rho$ and velocity dispersion~$v$ (equivalently, temperature). The second quantity is the relaxation time,~$t_{\rm relax}$, which fixes the 
thermal conductivity. The relaxation time is the characteristic time for DM to reach equilibrium either through self-interactions or interactions with other sectors,
such as baryons. The third quantity is the heating rate,~$\dot{\cal E}$, which is determined by the microphysics of DM-baryon interactions. 

Remarkably, the set of hydrodynamical equations is invariant under a one-parameter anisotropic space-time scaling transformation,~$\vec{x}\rightarrow \lambda \vec{x}$,~$t\rightarrow \lambda^z t$, for any~$z$, provided that the DM pressure, relaxation time and heating rate transform suitably. We take this as a powerful hint to fix the parametric dependence of each quantity. Starting with the equation of state, it turns out that the ideal gas form
\be
P = \rho v^2 
\label{ideal gas intro}
\ee
is invariant for any~$z$. What makes the ideal gas equation of state particularly appealing is its universality. It is valid as long as DM is sufficiently dilute, in the sense that the average inter-particle separation is large compared to the mean free path.  

The scaling symmetry requires that the relaxation time transform as~$t_{\rm relax}\rightarrow \lambda^z t_{\rm relax}$.
A natural choice in galactic dynamics which satisfies the desired scaling is the Jeans time, 
\be
t_{\rm relax} \sim \frac{1}{\sqrt{G\rho}}\,.
\label{t relax fix intro}
\ee
We will show that this choice allows us to reproduce the MDAR.

The final ingredient is the heating rate. To fix its form, we assume that the heating rate explicitly breaks scaling invariance for any~$z$ except~$z = 1/2$. This choice is empirically motivated by the BTFR, since the relevant ratio~$V_{\rm flat}^4/M_{\rm b}$ is invariant under the~$z = 1/2$ transformation. We will argue in Sec.~\ref{heat scaling} that this scaling, together with physically-plausible assumptions, fixes the dependence of the heating rate to
\be
\frac{\dot{{\cal E}}}{m} \sim a_0 v \frac{\rho_{\rm b}}{\rho} \,.
\label{heat rate intro}
\ee
The proportionality constant, which has units of acceleration, has been fixed empirically to match the MDAR characteristic acceleration scale~$a_0$. This scale must somehow emerge from the microphysics of DM-baryon interactions. 

Once the equation of state, relaxation time and heating rate are fixed, we will show that in the DM-dominated regime our equations enjoy a larger, approximate symmetry. Namely, the circular velocity curves~$V_1(R)$ and~$V_2(R)$ of two DM-dominated exponential disks with different scale lengths~$L_1$ and~$L_2$ and different total baryonic masses~$M_{{\rm b},1}$ and~$M_{{\rm b},2}$ must be related by:
\begin{equation}
	V_2(R) = \left(\frac{M_{\rm b,2}}{M_{\rm b,1}} \right)^{1/4} V_1\left(\tfrac{L_1}{L_2}R\right) \,.
\end{equation}
This encodes both the BTFR and the CSDR, at the root of the diversity of rotation curves.

We will then explore in more details in Sec.~\ref{MDAR broken scale} how Eqs.~\eqref{ideal gas intro}--\eqref{heat rate intro} are sufficient
ingredients to reproduce the MDAR. Specifically, we begin in Sec.~\ref{pseudo iso} by recalling how a cored pseudo-isothermal profile can, for suitable choice of its central density and core radius, reproduce the MDAR. Our working assumption, therefore, is that DM halos,
through DM self-interactions and baryon-DM energy exchange, reach a cored pseudo-isothermal profile in the region enclosing the galactic disk.

By focusing on static, equilibrium configurations, we proceed in Sec.~\ref{flat part BTFR} to show that the cored pseudo-isothermal profile, with suitable parameters to reproduce the MDAR, is a solution to our hydrodynamical equations. Specifically, in the flat part of the rotation curve the rotational velocity asymptotes to
\be
V_{\rm flat}^4 \sim a_0 GM_{\rm b} \log \frac{R_0}{r}\,.
\label{BTFRlog intro}
\ee
The prefactor matches the parametric dependence of the BTFR. Unfortunately within the static analysis we are unable to determine the arbitrary
radius~$R_0$ (which must be larger than the galaxy) or its scatter. Meanwhile, in the central region of galaxies, we show in Sec.~\ref{central HSB} that, for high-surface brightness (HSB) galaxies which are baryon-dominated near the center, the DM profile reproduces the CSDR with the behavior of the `simple' interpolating function of MOND~\cite{Famaey:2005fd}. In Sec.~\ref{approach to equi} we go beyond the equilibrium treatment and study the time-dependent approach to equilibrium, considering only average quantities suitable for the DM-dominated regime. This allows us to derive a particular combination of the DM velocity dispersion and surface density, which matches the combination of BTFR and CSDR. Therefore, if one takes the BTFR as a given (per the equilibrium analysis), this constraint yields the central density relation naturally for DM-dominated galaxies. 

We move on in Sec.~\ref{cosmo} to analyze the astrophysical and cosmological implications of our model. The form of the 
heating rate~\eqref{heat rate intro} allows us to derive very general results, irrespective of the underlying microphysical model.
The only assumption is that whatever DM-baryon interactions are at the root of this heat exchange still apply in the astrophysical/cosmological
context of interest. For this purpose, the inverse-density dependence of~$\dot{{\cal E}}/m$ is a welcome feature phenomenologically. It
implies a suppressed heat exchange in the early universe, allowing us to comfortably satisfy constraints from the CMB and the large
scale structure. Intriguingly, as shown in Sec.~\ref{edges sec} the heat exchange between DM and baryons, which acts to cool the
neutral gas prior to the Cosmic Dawn, provides a possible explanation to the anomalous EDGES signal at~$z\simeq 17$. This is unlike other DM-baryon
explanations of the EDGES excess, such as millicharged DM, which typically run afoul of CMB constraints~\cite{Kovetz:2018zan,Creque-Sarbinowski:2019mcm}.

It remains to construct a full-fledged model of particle physics that realizes the desired interactions. In the Conclusions section (Sec.~\ref{conclusions}) we will
discuss various promising avenues for model building to be pursued elsewhere.

\section{The MDAR and galactic scaling relations}
\label{MDAR review}

Since the MDAR (or MOND-like phenomenology) is an empirical fact about rotationally-supported galaxies, the scaling relations it implies must emerge in any phenomenologically-viable DM model. To set the stage, we begin with a brief review of the galactic scaling relations of interest. 

The MDAR is a relation between the total gravitational field~$g$ and the Newtonian acceleration~$g_{\rm b}$ generated by the observed distribution of baryons~\cite{McGaugh:2016leg}:
\be
g = \left\{\begin{array}{cl}
g_{\rm b} & \hspace{0.5cm} g_{\rm b}\gg a_0 \\
\sqrt{a_0g_{\rm b}} & \hspace{0.5cm} g_{\rm b}\ll a_0 \,,
\end{array}\right.
\label{MDAR}
\ee
where~$a_0 \simeq 10^{-10}\,{\rm m}/{\rm s}^2$. Numerically, this characteristic acceleration coincides with the Hubble scale~$a_0 \simeq \frac{1}{6} cH_0$.
The DM interpretation of the MDAR is that DM should only dominate when the baryonic acceleration drops below~$a_0$, and furthermore the effect of DM in this regime should be such that~$g\simeq \sqrt{a_0g_{\rm b}}$. 

An immediate corollary of the MDAR is the BTFR~\cite{LelliBTFR}. At large distances outside the baryon distribution, the baryonic acceleration can be approximated by~$g_{\rm b} \simeq G M_{\rm b}/r^2$, where~$M_{\rm b}$ is the total baryonic mass. Furthermore, in this regime the DM-dominated relation~$g\simeq \sqrt{a_0g_{\rm b}}$ applies. Substituting~$g = V^2_{\rm flat}/r$, where~$V_{\rm flat}$ is the rotational velocity, we obtain
\be
V_{\rm flat}^4 = a_0 G M_{\rm b}\,.
\label{BTFR}
\ee
Thus the MDAR implies the BTFR in the flat part of rotation curves, but the fact that galaxies obey the BTFR does not imply that they will obey the MDAR in the rising parts of rotation curves. The fact that they observationally do is at the root of the diversity of rotation curve shapes problem~\cite{Oman:2015xda,Ghari:2018jkc}. 

The diversity of shapes is related to the central surface density relation~\citep[CSDR,][]{Lelli:2016uea}, which is another consequence of the MDAR:
\be
\Sigma(0) = \left\{\begin{array}{cl}
\Sigma_{\rm b}(0) & \hspace{0.5cm} \Sigma_{\rm b}(0)\gg \tfrac{a_0}{G} \\
\sqrt{\frac{2}{\pi} \frac{a_0}{G} \Sigma_{\rm b}(0) } & \hspace{0.5cm} \Sigma_{\rm b}(0)\ll \tfrac{a_0}{G} \,,
\end{array}\right.
\label{surface relation}
\ee
where the central dynamical surface density~$\Sigma(0)=\int_{-\infty}^\infty {\rm d}z\, \rho(\vec{x})$, with~$z$ denoting the coordinate transverse to the disk, can be evaluated from the rotation curve. Similarly, the baryonic surface density is~$\Sigma_{\rm b} = \int_{-\infty}^\infty {\rm d}z\, \rho_{\rm b}(\vec{x})$. The dynamical surface density~$\Sigma$ is the sum of~$\Sigma_{\rm b}$ and the DM central surface density,~$\Sigma_{\rm DM}$. For a spherically-symmetric DM profile, the latter is defined by
\be
\Sigma_{\rm DM} = 2\int_0^\infty {\rm d}r\, \rho(r)\,. 
\label{SigmaDM}
\ee 
High-surface brightness (HSB) galaxies correspond to~$\Sigma_{\rm b}\gg a_0/G$ and are baryon-dominated in the central region. Low-surface brightness (LSB) galaxies have~$\Sigma_{\rm b}\ll a_0/G$ and are DM-dominated everywhere. 

LSB galaxies are particularly interesting because they imply a scaling symmetry, which is at the root of the MOND paradigm~\cite{Milgrom:1983ca,Milgrom:1983pn, Milgrom:2008cs}. Indeed the idea of MOND is that below the acceleration scale~$a_0$, corresponding to the DM-dominated regime, dynamics are invariant under the space-time scaling 
\be
\vec{x}\rightarrow \lambda \vec{x}\,;\qquad t\rightarrow \lambda  t \,.
\ee
This implies, in particular, that, two LSB exponential disks of same total mass~$M_{\rm b}$ but different scale-lengths~$L_1$ and~$L_2$, will have identical rotation curves
expressed in scale-length units. More generally, combining this with the BTFR, the circular velocities~$V_1$ and~$V_2$ of two LSB disks should be related by
\begin{equation}
	V_2(R) = \left(\frac{M_{\rm b,2}}{M_{\rm b,1}} \right)^{1/4} V_1\left(\tfrac{L_1}{L_2} R\right)\,,
\label{eq:LSBscaling}
\end{equation}
where~$R$ is the axisymmetric radius within the galactic plane of each galaxy.

One can think of the above scaling relations as follows. The BTFR~\eqref{BTFR} is a {\it global} constraint, relating the asymptotic rotational velocity to the total baryonic mass at large~$R$. The CSDR~\eqref{surface relation} constrains the total and baryonic central surface densities as~$R\rightarrow 0$. For DM-dominated LSB galaxies, these two scaling relations can be summarized by the scale invariant equation (\ref{eq:LSBscaling}). More generally, all these scaling relations can be summarized by the MDAR~\eqref{MDAR}, which is a {\it local} relation between the baryonic and DM gravitational accelerations valid at every point in the galaxy.

\section{Baryon-Interacting Dark Matter}
\label{setup}

We begin with a brief review of the general framework laid out in~\cite{Famaey:2017xou}. The starting point is a generalization of the usual collisionless Boltzmann equation for DM to a {\it Boltzmann transport equation}, which includes a collisional integral encoding interactions between DM particles and baryons. For simplicity, we will restrict our attention to the zeroth, first and second velocity moments of this equation, which respectively enforce mass, momentum and energy conservation:
\begin{subequations} \label{time dependent equations}
\bea
\label{continuity}
& \displaystyle \frac{\partial \rho}{\partial t} + \vec{\nabla}\cdot\left(\rho \vec{u}\right) = 0\,; & \\
\label{momentum}
& \displaystyle  \left(\frac{\partial}{\partial t}  + \vec{u}\cdot\vec{\nabla}\right)u^i   + \frac{1}{\rho} \partial_jP^{ij} = g^i\,; &\\
\label{energy}
&\displaystyle  \frac{3}{2} \left(\frac{\partial}{\partial t}  + \vec{u}\cdot\vec{\nabla}\right) \frac{T}{m} +  \frac{1}{\rho} P^{ij}  \partial_i u_j + \frac{1}{\rho} \vec{\nabla}\cdot \vec q = \frac{\dot{{\cal E}}}{m}\,. &
\eea
\end{subequations}
Here,~$\vec{u} \equiv \langle \vec{v} \rangle$ is the bulk DM velocity,~$P^{ij} \equiv \rho \left\langle \left(v^i - u^i\right)\left(v^j - u^j\right)\right\rangle$ is the pressure tensor,~$T \equiv \frac{m}{3} \langle |\vec{v}-\vec{u}|^2\rangle$ is the local DM temperature, and~$\vec{q} \equiv \frac{1}{2}\rho \langle (\vec{v}-\vec{u})|\vec{v}-\vec{u}|^2\rangle$ is the heat flux. The local heating rate~$\dot{{\cal E}}$ is due to interactions with baryons. The (total) gravitational acceleration~$\vec{g}$ is determined as usual by the Poisson equation
\be
\vec{\nabla}\cdot \vec{g}= -4\pi G\left(\rho + \rho_{\rm b}\right)\,.
\label{poisson}
\ee
The baryon mass density~$\rho_{\rm b}(\vec{x})$ will be treated as an input specified by observations. Moreover, in what follows we will be interested in velocity distributions that are approximately isotropic, in which case
\begin{equation}
	P_{ij} \simeq P \delta_{ij} \qquad \text{valid for}~~| \vec u| \ll  v\,, 
\end{equation}
where we have introduced the one-dimensional velocity dispersion~$v=\sqrt{T/m}$.

\subsection{General scaling symmetry}

Having reviewed the framework of~\cite{Famaey:2017xou}, let us discuss the scaling properties of the above equations. Setting~$\dot{{\cal E}} = 0$ temporarily, notice that~\eqref{time dependent equations} and~\eqref{poisson} are invariant under the anisotropic space-time scaling transformation 
\be
\vec{x}\rightarrow \lambda \vec{x}\,;\qquad t\rightarrow \lambda^z t \,,
\label{scaling z}
\ee
valid for {\it arbitrary}~$z$, with the various quantities transforming as\footnote{Note that this scaling symmetry is different than the one considered in~\cite{Famaey:2017xou} because~$\rho_{\rm b}$ transforms differently. They agree only for~$z = 1/2$.} 
\bea
\nonumber
v & \rightarrow & \lambda^{1-z} v  \,;\\
\nonumber
\vec{u} & \rightarrow & \lambda^{1-z} \vec{u}  \,;\\
\nonumber
\vec{g} & \rightarrow & \lambda^{1-2z}\vec{g} \,;\\
\label{z symmetry}
\rho & \rightarrow & \lambda^{-2z} \rho  \,;\\
\nonumber
\rho_{\rm b} & \rightarrow & \lambda^{-2z} \rho_{\rm b}  \,;\\
\nonumber
P^{ij} & \rightarrow & \lambda^{2-4z} P^{ij}  \,;\\
\vec{q} & \rightarrow & \lambda^{3-5z} \vec{q}\,.
\nonumber 
\eea
Notice that the transformation laws for~$P^{ij}$ and~$\vec{q}$ are compatible with their definition in terms of~$\rho$,~$\vec{v}$ and~$\vec{u}$. The above is a symmetry of the collisionless equations. In order for it to survive as a symmetry of the collisional equations ({\it i.e.}, with
non-zero~$\dot{{\cal E}}$), the heating rate must transform as 
\be
\frac{\dot{{\cal E}}}{m} \rightarrow \lambda^{2-3z} \frac{\dot{{\cal E}}}{m}\,.
\label{rate transform}
\ee

The transformation rules~\eqref{z symmetry} and~\eqref{rate transform} could at first glance be dismissed as a trivial consequence of dimensional analysis, with units of length and time kept separate due to the non-relativistic nature of our system. This becomes more manifest by rescaling~$\rho$,~$\rho_\text{b}$,~$P$,~$\vec q$ and~$\dot{{\cal E}}$ in Eqs.~\eqref{z symmetry} and~\eqref{rate transform} by a factor of~$G$---a procedure that does not affect Eqs.~\eqref{time dependent equations}. Nevertheless, in what follows we will demand that this scaling is actually an emergent symmetry of the DM sector and its interactions with baryons, at least for a specific value of~$z$. This requirement, together with some physically-motivated assumptions, will place stringent constraints on the DM equation of state, the heat flux, and the heating rate.

\subsection{DM equation of state}

In order to solve Eqs.~\eqref{time dependent equations} one must specify, among other things, an equation of state for DM, which for our purposes will be a relation of the form~$P = P (\rho, v)$. The explicit form of such a relation depends on the microscopic details of the DM sector. The requirement that the equation of state be scale invariant for some particular value of~$z$ places a nontrivial constraint on its functional form.

Remarkably, there is a very general assumption one can make to obtain an equation of state that is scale invariant {\it for any}~$z$. Namely, we assume that DM is sufficiently dilute, in the sense that~$n \lambda^3 \ll 1$, where~$n = \tfrac{\rho}{m}$ is the number density of DM particles, and~$\lambda = \tfrac{1}{mv}$ their mean thermal wavelength. In this regime one can perform a virial expansion of the DM equation of state, which at lowest order generically reduces to that for an ideal gas:
\be
P = \rho v^2\, .
\label{ideal gas}
\ee
It is easy to check that this relation is the {\it only} equation of state that is invariant under the symmetry transformations~\eqref{z symmetry} for arbitrary~$z$.

\subsection{Heat flux and relaxation time}

In the limit where deviations from thermal equilibrium are small,\footnote{To be more precise, in the spherically symmetric case we will consider later on, Fourier's law is valid provided~$\left\vert \tfrac{{\rm d} \log v^2}{{\rm d} \log r}\right\vert \ll 1$.} Fourier's law provides us with an approximate yet explicit expression for the heat flux~$\vec q$: 
\be
\vec q \simeq - \kappa m \vec \nabla v^2 \,,	
\label{Fourierlaw}
\ee
where~$\kappa$ is the thermal conductivity,
\be
\kappa = \mathcal{O}(1) \frac{\rho \, v^2t_{\rm relax}}{m} \ ,
\label{conduct}
\ee
and~$t_{\rm relax}$ denotes the relaxation time. This parameter can be thought of as the characteristic time for DM to reach equilibrium due to self-interactions or interactions with other sectors, {\it e.g.} with baryons.

The scaling transformations~\eqref{z symmetry} immediately imply that ~$t_{\rm relax}$ must transform as a time scale:
\be
t_{\rm relax}\rightarrow \lambda^z \, t_{\rm relax}\,. \label{relaxation time scaling}
\ee
Once again one might be tempted to attribute this scaling to dimensional analysis and therefore conclude that it is devoid of any physical significance. However, a generic relaxation mechanism will emphatically {\it not} give rise to a~$t_{\rm relax}$ with this scaling property for arbitrary values of~$z$. Imagine for instance that DM reaches thermal equilibrium due to self-interactions. The cross section for such processes will generically have a velocity dependence of the form~$\sigma = \sigma_0 (c/v)^\alpha$ for a fixed~$\alpha$, and with~$\sigma_0$ a constant built out of microscopic scales and couplings. The relaxation time is in turn the inverse of the self-interaction rate~$\sigma n v$, {\it i.e.},~$t_{\rm relax} = \tfrac{m (v/c)^\alpha}{\sigma_0 \rho v}$. We conclude therefore that in this scenario~$t_{\rm relax} \to \lambda^{(3-\alpha)z-1+\alpha} t_{\rm relax}$, which agrees with~\eqref{relaxation time scaling} only for one particular value of~$z$, namely~$z = \tfrac{1-\alpha}{2-\alpha}$.

More broadly, one should keep in mind that multiple relaxation mechanisms might be at play over different characteristic time scales, in which case the relaxation time should be the shortest of such scales. Given that there is currently no direct evidence for sizable DM self-interactions, it is plausible that the associated time scale could be longer than the dynamical time in galaxies. It is then important to consider the possibility of other relaxation mechanisms. This naturally suggests another time scale, which interestingly scales like~\eqref{relaxation time scaling} for any~$z$---the Jeans time~$\tfrac{1}{\sqrt{G\rho}}$. A possible mechanism giving rise to such a relaxation time was discussed for instance in~\cite{Famaey:2017xou}. 

Indeed, we will see below that, in order to reproduce the MDAR, the relaxation time must indeed be proportional to the Jeans time, {\it i.e.},
\be
t_{\rm relax} = \frac{\mathcal{O}(1) }{\sqrt{G\rho}}\,.
\label{t relax fix}
\ee
 In the flat part of the rotation curve, where~$\rho(r) \simeq \frac{v^2}{2\pi G r^2}$, this reduces to~$t_{\rm relax} \sim \tfrac{r}{v}$.
Combining this expression with the one for the thermal conductivity in Eq.~\eqref{conduct}, we obtain
\be
\kappa m= {\cal N} \sqrt{\frac{\rho}{G}} \, v^2\, ,
\label{conduct final}
\ee
where~${\cal N}$ is some~${\cal O}(1)$ constant.

\subsection{Heating rate }
\label{heat scaling}

By working in the dilute limit and assuming that~$t_{\rm relax}$ is determined by the Jeans time, we have been able to ``kick the can down the road'' and preserve scale invariance without committing to any particular value of~$z$. In order to write down an explicit expression for the heating rate, we will now have to fix~$z$. 

To this end we will use the BTFR as an observational guiding principle. The fact that the ratio~$V_{\rm flat}^4 / M_{\rm b}$ appears to be a universal constant in rotationally-supported galaxies suggests that this quantity should not transform under our scaling symmetry. This will be the case only if the scaling exponent takes the value
\be
z =1/2\,.
\ee
We henceforth assume that our heating rate explicitly breaks scale invariance for any~$z$ down to scale invariance for~$z=1/2$ only. 

We will now show, based on plausible physical assumptions, that the~$z=1/2$ scaling symmetry 
\be
\frac{\dot{{\cal E}}}{m} \rightarrow \lambda^{1/2} \frac{\dot{{\cal E}}}{m}\, , 
\label{rate transform z=1/2}
\ee
fixes the parametric dependence of the heating rate~$\dot{\cal E}/m$ due to DM-baryon interactions. On physical grounds, we expect~$\dot {\cal E}/m$ to depend on~$\rho, \rho_{\rm b}$, both of which transform as~$\rho_{\rm b}, \rho\rightarrow \lambda^{-1} \rho_{\rm b}, \rho$, as well as the velocity of DM and baryon components. In rotationally-supported galaxies it is reasonable to neglect the DM bulk velocity relative to its velocity dispersion,~$|\vec{u}| \ll v$. Indeed, in most of our analysis we will focus on equilibrium situations and ignore the spin of the halo. We will assume the opposite for baryons, 
$v_{\rm b}\ll |\vec{V}_{\rm b}|$, which is also justified in disk galaxies. This leaves us with two velocity variables,~$v$ and~$V_{\rm b}$. These two are comparable in the flat part of rotation curves, whereas~$V_{\rm b} \ll v$ in the central region of galaxies. To simplify the discussion, we shall only keep track of the dependence on~$v$, keeping in mind that~$\dot {\cal E}/m$ more generally will depend on both~$v$ and~$V_{\rm b}$. 

Given the transformation law~$v \to \lambda^{1/2} v$, the most general form for the heating rate compatible with~\eqref{rate transform z=1/2} is
\begin{equation}
	\frac{\dot{{\cal E}}}{m} = v \, F \left(\tfrac{\rho_{\rm b}}{\rho}, \tfrac{v^2}{\rho}\right)\,.
\end{equation}
In order to fix completely the form of~$\dot{{\cal E}}$, we will make two additional assumptions. First, since in our scenario DM heats up due to interactions with baryons, it is natural to assume that it is an extensive quantity as a function of the number of baryons. In other words, the heating rate should be linear in~$\rho_{\rm b}$:
\begin{equation}
	\frac{\dot{{\cal E}}}{m} = v \, \frac{\rho_{\rm b}}{\rho} f \left(\tfrac{v^2}{\rho}\right)\,.
	\label{heating rate extensive}
\end{equation}
From a model-building perspective, this is certainly the simplest possibility. This is arguably also the most reasonable behavior one can have in the DM dominate regime~$\rho_{\rm b} / \rho \ll 1$. We will assume however that Eq.~\eqref{heating rate extensive} holds more generally.

Notice that~$f$ has dimensions of acceleration. Therefore, the second assumption we will make is that the~$f$ is approximately constant, and of order the characteristic acceleration scale~$a_0$ appearing in the MDAR.  Thus the heating rate is fixed to be
\be
\frac{\dot{{\cal E}}}{m} = C a_0 v \frac{\rho_{\rm b}}{\rho} \,,
\label{heat rate final}
\ee
where~$C$ is another constant. For concreteness we will assume~$C \sim {\cal O}(10^{-1})$, which offered a good fit to rotation curves in the cooling case~\cite{Famaey:2017xou}.
The assumption that~$f$ is of order~$a_0$ is also quite natural from a phenomenological viewpoint, given that we are trying to reproduce a result such as the MDAR which features a characteristic acceleration scale. At the same time, the obvious downside of treating~$a_0$ as a fundamental scale is that it is unclear why it should numerically coincide with a cosmological acceleration scale. We will assume that this ``coincidence'' is resolved by a different mechanism that operates over much longer, cosmological time scales, such that~$a_0$ can be treated as a constant parameter for our purposes. This appears to be well supported by current observations~\cite{Lelli:2018swe}. It is also worth noting that the inverse density dependence in~\eqref{heat rate final} is helpful for the phenomenological viability of the mechanism. As we will see in Sec.~\ref{cosmo}, it suppresses the heating rate in high-density environments, such as the early universe. 

Finally, a brief word about the sign of~$C$, which determines whether DM is cooled ($\dot{{\cal E}} < 0$) or heated ($\dot{{\cal E}} > 0$) by baryons.
Whereas~\cite{Famaey:2017xou} primarily studied the cooling case for concreteness, here we focus exclusively on the heating case. This is {\it a priori} more desirable,
since DM heating can transform the cusps into cores in the central regions of galaxy halos. Moreover, the opposite case of DM cooling can lead to flattened halos, or too prominent dark disks,
once the halos have an initial spin. These unwanted features are absent with DM heating. Finally, we will argue in Sec.~\ref{approach to equi} that with heating it is possible to derive a combination of the BTFR and CSDR by studying the dynamical approach to equilibrium. 

\subsection{Deep-MOND scaling as an approximate enhanced symmetry}
\label{sec: enhanced symmetry}

To summarize, given our expressions for the equations of state, the heat flux and the heating rate, Eqs.~\eqref{time dependent equations} reduce to:
\begin{subequations} \label{time dependent final equations}
\bea
\label{final continuity}
& \displaystyle \frac{\partial \rho}{\partial t} + \vec{\nabla}\cdot\left(\rho \, \vec{u}\right) = 0\,; & \\
\label{final momentum}
& \displaystyle  \left(\frac{\partial}{\partial t}  + \vec{u}\cdot\vec{\nabla}\right)\vec{u}   +  \frac{1}{\rho} \vec{\nabla} \left(\rho v^2\right) = \vec{g}\,; &\\
\label{final energy}
&\displaystyle  \frac{3}{2} \left(\frac{\partial}{\partial t}  + \vec{u}\cdot\vec{\nabla}\right) v^2 +  v^2  \, \vec{\nabla} \cdot \vec u - \frac{1}{\rho} \vec{\nabla}\cdot\left(  {\cal N} \sqrt{\frac{\rho}{G}} \, v^2 \vec \nabla v^2 \right) = C a_0 v \, \frac{\rho_{\rm b}}{\rho}\,; & \\
&\displaystyle \vec{\nabla}\cdot \vec{g}= -4\pi G\left(\rho + \rho_{\rm b}\right)\,.
\label{final poisson} & 
\eea
\end{subequations}
As discussed previously, these equations are invariant under the scaling transformations~\eqref{scaling z} and~\eqref{z symmetry} with~$z = 1/2$. 

In fact, in the DM-dominated regime, where~$\rho_{\rm b}$ can be neglected compared to~$\rho$ in the Poisson equation~\eqref{final poisson},\footnote{Notice that in this limit one cannot necessarily neglect the righthand side of Eq.~\eqref{final energy}. For instance, for equilibrium solutions the right-hand side is exactly equal to the last term on the left-hand side, and is therefore not negligible.} our equations enjoy a larger, approximate symmetry under the rescaling
\begin{eqnarray}
\nonumber
\vec x & \rightarrow & \lambda \, \vec x  \,;\\
\nonumber
t & \rightarrow & \lambda^y \, t  \,;\\
\nonumber
v & \rightarrow & \lambda^{1-y} v  \,;\\
\label{y symmetry}
\vec{u} & \rightarrow & \lambda^{1-y} \vec{u}  \,;\\
\nonumber
\vec{g} & \rightarrow & \lambda^{1-2y}\vec{g} \,;\\
\nonumber
\rho & \rightarrow & \lambda^{-2y} \rho  \,;\\
\nonumber
\rho_{\rm b} & \rightarrow & \lambda^{1-4y} \rho_{\rm b}  \,,
\end{eqnarray}
for an arbitrary~$y$~\cite{Famaey:2017xou}. These transformations  reduce to our original~$z = 1/2$ scale symmetry for~$y=1/2$, but for other values of~$y$ they represent a new type of symmetry that is only approximately valid in DM-dominated regions. 

Despite its approximate validity, this enhanced symmetry has interesting observational consequences. Imagine that a galaxy with scale length~$L_1$, total baryonic mass~$M_{\rm b,1}$ and rotation curve~$\vec{V}_1$ is a solution to our equations. It immediately follows that our equations must also admit a solution with~$L_2$,~$M_{\rm b,2}$ and~$\vec{V}_2$ given by
\begin{equation}
	L_2 = \lambda L_1\,; \qquad \qquad M_{\rm b,2} = \lambda^{4-4y} M_{\rm b,1}\,; \qquad \qquad \vec{V}_2(\lambda \vec{x}) = \lambda^{1-y} \vec{V}_1\left(\vec{x}\right) \,. 
	\label{generalized rotation curve scaling}
\end{equation} 
This is equivalent to the statement that the rotation curves of two galaxies with different scale lengths and different total baryonic masses must be related as follows:
\begin{equation}
	\vec{V}_2(\vec{x}) = \left(\frac{M_{\rm b,2}}{M_{\rm b,1}} \right)^{1/4} \vec{V}_1\left(\tfrac{L_1}{L_2}\vec{x}\right) \,,
\end{equation}
which precisely matches~\eqref{eq:LSBscaling}. 

In the particular case of~$y=1$, the scaling transformations~\eqref{y symmetry} reduce to the ``relativistic'' deep-MOND scaling law~\cite{Milgrom:2008cs}, and the result~\eqref{generalized rotation curve scaling} becomes particularly simple: two galaxies with the same total baryonic mass but different scale lengths~$L_1$ and~$L_2$ have rotation curves related by~$\vec{V}_2(\vec{x}) = \vec{V}_1\left(\tfrac{L_1}{L_2}\vec{x}\right)$. This behavior appears to be supported by observations~\cite{Swaters:2009by}.

\section{MDAR as Spontaneous Breaking of Scale Invariance}
\label{MDAR broken scale}

As shown above, the scaling of our equations implies that, in the DM-dominated regime, the baryonic mass-asymptotic velocity scaling should follow the BTFR scaling,~$M_{\rm b} \propto V_{\rm flat}^4$. Regarding the normalization of the BTFR, it is known that if one starts from abundance matching with NFW halos, one typically reproduces the correct zero-point of the relation in the baryonic mass range~$\sim 10^{10} {\rm M}_\odot$ to~$\sim 10^{11} {\rm M}_\odot$, albeit with too large scatter~\cite{Desmond:2017}. The curvature of the predicted BTFR then implies too large~$V_{\rm flat}$ (or too large enclosed DM mass) at the low-mass end, still with too large scatter. Given that we are starting from the right normalization in the intermediate-mass regime, one would expect that our heating mechanism expels DM out of the baryonic disk region of low-mass disk galaxies, thereby bringing~$V_{\rm flat}$ down to follow the~$M_{\rm b} \propto V_{\rm flat}^4$ scaling with the zero-point set by intermediate-mass galaxies.

In order to make more concrete analytic predictions hereafter, we will now assume that, through their own self-interactions together with the baryon-DM energy exchange mechanism, DM halos reach a cored pseudo-isothermal profile in the region where the baryonic disk is sitting. In this Section we will demonstrate that the set of equations~\eqref{time dependent final equations} is fully consistent with such a cored pseudo-isothermal profile, with parameters that reproduce the MDAR. 

\subsection{Cored pseudo-isothermal profile}
\label{pseudo iso}

Let us now first show how the cored pseudo-isothermal profile parameters should be arranged to reproduce the MDAR. The profile has the following form:
\be
\rho(r) = \frac{\rho_0}{1 + \left(\frac{r}{r_{\rm c}}\right)^2}\,.
\label{COP}
\ee
Thus it is specified by two parameters: the central density,~$\rho_0$, and the core radius,~$r_{\rm c}$. Equivalently, the
core radius can be traded for the (asymptotic) velocity dispersion, denoted by~$v_\infty$, using
\be
r_{\rm c} = \frac{v_\infty}{\sqrt{2\pi G\rho_0}}\,.
\label{R v rho0}
\ee
Note that~$v_\infty$ is defined at infinity because the velocity dispersion profile we are considering is {\it not} strictly isothermal.

The ability of such cored pseudo-isothermal profile to fit galactic rotation curves has been well-studied, {\it e.g.},~\cite{Jimenez:2002vy}. 
Consider first the large distance~$r\gg r_{\rm c}$ regime:
\be
\rho(r\gg r_{\rm c}) \simeq \frac{\rho_0 r_{\rm c}^2}{r^2} = \frac{v^2_\infty}{2\pi G r^2} \,.
\label{rho large r}
\ee
This implies a flat rotation curve with~$V_{\rm flat} = \sqrt{2} v_\infty$. Hence DM dominates in this regime, and the assumption of
spherical symmetry is justified. To match the BTFR~\eqref{BTFR}, the velocity dispersion must be related to the total baryonic mass via
\be
v^4_\infty = \frac{1}{4} a_0GM_{\rm b}\,.
\label{v cons}
\ee
This fixes one parameter of the cored pseudo-isothermal profile~\eqref{COP}, which thus simplifies to
\be
\rho(r) = \frac{1}{4\pi G} \frac{\sqrt{a_0GM_{\rm b}}}{r_{\rm c}^2 + r^2}\,.
\label{COP2}
\ee
 
The second parameter can be fixed by the CSDR~\eqref{surface relation}.
For the cored pseudo-isothermal profile,~\eqref{SigmaDM} gives
\be
\Sigma_{\rm DM} = \pi \rho_0 r_{\rm c}\,.
\label{sigma DM}
\ee
To proceed, we must distinguish between LSB galaxies, which are DM-dominated everywhere, and HSB galaxies, where baryons
dominate in the central region. For LSB galaxies ($\Sigma_{\rm b}\ll a_0/G$),~\eqref{surface relation} implies 
\be
\rho_0 r_{\rm c} = \frac{2}{\pi} \sqrt{\frac{a_0}{2\pi G}\Sigma_{\rm b}(0)}\,.
\ee
Combined with~\eqref{R v rho0} and the first constraint~\eqref{v cons}, we can solve for the core radius of LSB galaxies:
\be
r_{\rm c} = \frac{1}{4} \sqrt{\frac{\pi M_{\rm b}}{2\Sigma_{\rm b}(0)}}\qquad (\text{LSB galaxies})\,.
\label{R LSB}
\ee
For HSB galaxies ($\Sigma_{\rm b}\gg a_0/G$), on the other hand, the CSDR~\eqref{surface relation} does not directly constrain~$\Sigma_{\rm DM}$.
The answer depends on the assumed functional form for the MDAR. (In the MOND parlance, this reflects the freedom in choosing the interpolating function.)

From a symmetry perspective, the cored pseudo-isothermal profile spontaneously breaks the~$z=1/2$ scaling symmetry by introducing an explicit scale,~$r_{\rm c}$ (or equivalently,~$\rho_0$).
Notice, however, that the scaling symmetry is restored in the flat part of the rotation curve ({\it i.e.},~$r \gg r_{\rm c}$). Indeed, in this region~$\rho(r)$ approximates a singular isothermal
profile~\eqref{rho large r}, which transforms covariantly for any~$z$:
\be
\rho(r) \simeq \frac{v^2_\infty}{2\pi G r^2} \rightarrow  \lambda^{-2z} \rho(r)\,.
\ee
The spontaneous symmetry breaking scale~$r_{\rm c}$ (as well as~$v_\infty$) will be fixed through other sources of spontaneous breaking, namely baryons.  

\subsection{Flat part of the rotation curve and the BTFR} 
\label{flat part BTFR}

We now show that a cored pseudo-isothermal profile, with suitable parameters to reproduce the MDAR, is a solution to the set of
equations~\eqref{time dependent final equations}.  We will primarily be interested in equilibrium solutions to these equations with
negligible DM halo spin. In this case, the DM bulk velocity can be set to zero, {\it i.e.},~$\vec{u} = 0$, and the continuity
equation~\eqref{final continuity} is trivially satisfied. Equations~\eqref{final momentum}--\eqref{final poisson} then reduce to
\begin{subequations} \label{time indep final equations}
\bea
\label{final momentum 2}
& \displaystyle  \vec{\nabla} \left(\rho v^2\right) = \rho \vec{g}\,; &\\
\label{final energy 2}
&\displaystyle  \vec{\nabla}\cdot\left( \sqrt{\frac{\rho}{G}} \, v^2 \vec \nabla v^2 \right) = -\frac{C}{{\cal N}} v a_0 \rho_{\rm b}\,; & \\
&\displaystyle \vec{\nabla}\cdot \vec{g}= -4\pi G\left(\rho + \rho_{\rm b}\right)\,.
\label{final poisson 2} & 
\eea
\end{subequations}

In the flat part of the rotation curve ($r \gg r_{\rm c}$), the gravitational field is dominated by DM ($\rho \gg \rho_{\rm b}$), and spherical symmetry is a good approximation.
The Jeans equation~\eqref{final momentum 2} and Poisson equation~\eqref{final poisson 2} are approximately solved by
\be
\rho(r) \simeq \frac{v^2(r)}{2\pi G r^2}\,,
\label{rho(r) v(r)}
\ee
where, as we will verify {\it a posteriori},~$v(r)$ is a slowly-varying function. Meanwhile, the velocity profile~$v(r)$ is determined by the heat
equation~\eqref{final energy 2}, which, upon assuming spherical symmetry and using~\eqref{rho(r) v(r)}, simplifies to
\be
\frac{1}{r^2} \frac{{\rm d}}{{\rm d}r} \left(v^4 r \frac{{\rm d}v}{{\rm d}r}\right) = -\sqrt{\frac{\pi}{2}}\frac{C}{{\cal N}} v a_0 G \rho_{\rm b}\,.
\label{heat spherical}
\ee
Approximating~$v$ as nearly constant on the right-hand side, this can be readily integrated once:
\be
r \frac{{\rm d} v^4}{{\rm d}r} = -\frac{1}{\sqrt{2\pi}} \frac{C}{{\cal N}}a_0 GM_{\rm b}\,.
\ee
In turn this implies
\be
v^4(r) = \frac{1}{\sqrt{2\pi}} \frac{C}{{\cal N}}a_0 GM_{\rm b} \log \frac{R_0}{r}\,,
\ee
where~$R_0$ is an arbitrary scale. Thus~$v$ only varies logarithmically, which justifies our assumption. 

Some remarks are in order. First, the logarithmic dependence of~$v(r)$ implies that scale invariance is not quite restored for~$r\gg r_{\rm c}$. Rather it is spontaneously
broken, analogously to the breaking of scale invariance by radiative corrections (as in Coleman-Weinberg~\cite{Coleman:1973jx}), with~$R_0$ playing the role of a dimensional
transmutation scale. Second, using the approximate relation~$V \simeq \sqrt{2} v$, the rotation curve is nearly flat with 
\be
V_{\rm flat}^4 \sim a_0 GM_{\rm b} \log \frac{R_0}{r}\,.
\label{BTFRlog final}
\ee
It is encouraging that the prefactor matches the parametric dependence of the BTFR~\eqref{BTFR}. Unfortunately within our static equilibrium analysis we are
not able to fix the scale~$R_0$, nor determine its scatter. To do so, we will need to go beyond the equilibrium treatment and analyze the dynamical evolution
towards equilibrium. This will be the focus of Sec.~\ref{approach to equi}.

\subsection{Cored region and the central density relation in HSB galaxies} 
\label{central HSB}

Consider the central region of galaxies ($r \ll r_{\rm c}$). In this region the DM density can be approximated as nearly constant,~$\rho\simeq \rho_0$,
hence~\eqref{final momentum 2} reduces to
\be
\vec{\nabla}v^2  \simeq \vec{g}\,.
\label{v^2 g}
\ee
The solution is~$v^2 = - \Phi  + \alpha v_\infty^2$, where~$\alpha$ is an~${\cal O}(1)$ constant. The precise value of this constant is irrelevant for us. The important point is that~$v^2$ approaches~$\sim v_\infty^2$ near the origin, while its gradient is fixed by the gravitational field. 

To make headway analytically, we imagine working sufficiently close to the center that the baryon distribution looks like an infinite disk but sufficiently far that the disk appears infinitely thin. In other words, we work in the regime~$L_{\rm z} \ll r \ll L$, where~$L_{\rm z}$ is the scale height and~$L$ the disk length of the baryon distribution. As a result, the baryon distribution is approximated by a surface density~$\Sigma_{\rm b}$:
\be
\rho_{\rm b} \simeq \Sigma_{\rm b} \delta(z)\,. 
\ee
For distances~$\ll L$, the surface density is nearly homogeneous and given by the central value,~$\Sigma_{\rm b}(0)$. 

With this approximation, the heat equation~\eqref{final energy 2} implies a discontinuity in the normal component of the heat flux, which by symmetry fixes its magnitude:
\be
\sqrt{\frac{\rho_0}{G}} v_{\infty} |\nabla_\perp v^2| =  \frac{C}{2{\cal N}} a_0 \Sigma_{\rm b}(0)\,.
\ee
Using~\eqref{R v rho0},~\eqref{sigma DM}, and~\eqref{v^2 g}, this implies
\be
\Sigma_{\rm DM}  g_\perp = \sqrt{\frac{\pi}{2}}\frac{C}{2{\cal N}} a_0 \Sigma_{\rm b}(0)\,.
\label{central reln}
\ee

The transverse component of the gravitational field is solved similarly by integrating Poisson's equation~\eqref{final poisson 2}.
For HSB galaxies, which are baryon-dominated near the center, this gives
\be
g_\perp^{\rm HSB} \simeq 2\pi G \Sigma_{\rm b}(0)\,.
\ee
It then follows from~\eqref{central reln} that
\be
\Sigma_{\rm DM} = \sqrt{\frac{\pi}{2}}\frac{C}{2{\cal N}}  \frac{a_0}{2\pi G}\qquad (\text{HSB galaxies})\,.
\label{Sigma HSB}
\ee
Thus our heat equation implies~$\Sigma_{\rm DM} \sim a_0/G$. This matches behavior of the `simple' interpolating function~\cite{Famaey:2005fd}, and is consistent with observations~\cite{Donato:2009ab}.

\subsection{Approach to equilibrium and central density relation in LSB galaxies}
\label{approach to equi}

Up to now our analysis has focused on static, equilibrium configurations. Within this framework, we were able to reproduce the parametric dependence of the BTFR, up to the
logarithm of a scale~$R_0$ whose magnitude and scatter remain undetermined. We were also able to derive the CSDR for HSB galaxies. 

By going beyond the equilibrium treatment and considering the approach to equilibrium, we will now show how the central density relation, which is at the
root of the problem of diversity of rotation curves, can be naturally reached by our DM-fluid interacting with baryons. Specifically, we will derive a constraint on a particular combination of the DM temperature and surface density, which matches the combination of BTFR and CSDR. Therefore, if one takes the BTFR as a given (per the equilibrium analysis), then
this constraint yields the central density relation naturally. 

We begin with a few general comments. In the standard~$\Lambda$CDM model, halo virialization is achieved through violent relaxation, a manifestly non-equilibrium
process that drives the DM distribution towards the attractor NFW profile within a few dynamical times. Our proposed DM-baryon
interactions offer another relaxation channel. These interactions have a characteristic time on the order of a dynamical time and thus ``compete" with violent
relaxation~\cite{Famaey:2017xou}. Therefore we do not expect our halos to necessarily reach a NFW profile early on. Crucially, since the interactions considered here tend to heat up DM, they can plausibly prevent the formation of cold central cusps and instead generate constant density cores, as needed in most LSB galaxy halos.  

A rigorous dynamical analysis to back this intuition would require numerical simulations, which is beyond the scope of this work. In what follows we offer a simple, back-of-the-envelope analysis of the time-dependent problem. Because the derivation ignores density and velocity gradients, and relies instead on average quantities, it can only reproduce the CSDR in the DM-dominated regime (valid for LSB galaxies). This is sufficient for our purposes, since we have already established the central density relation in HSB galaxies within the equilibrium treatment.

The starting point is our set of DM fluid equations~\eqref{time dependent final equations}. It is convenient to translate these equations in terms of the entropy density per DM particle,
given by the Sackur-Tetrode equation:
\be
s = \ln \left((2\pi)^{3/2} \frac{m^4v^3}{\rho}\right) + \frac{5}{2}\,.
\label{ST equation}
\ee
This allows us to eliminate~$v$ and express our equations~\eqref{time dependent final equations} in terms of~$\rho$,~$\vec{u}$ and~$s$. 
In what follows we will keep~$v$ around for simplicity, but it should be understood via~\eqref{ST equation} as an implicit function of~$\rho$ and~$s$.
It is straightforward to combine the continuity~\eqref{final continuity} and heat equation~\eqref{final energy} to obtain an equation for the
entropy density: 
\be
\left(\frac{\partial}{\partial t}  + \vec{u}\cdot\vec{\nabla}\right) s + \frac{1}{\rho v^2}\vec{\nabla}\cdot \vec{q} = \frac{\dot{{\cal E}}}{m} \,,
\label{entropy eqn}
\ee
with the heat flux expressed as
\be
\vec{q} = - \frac{2}{3} {\cal N} \sqrt{\frac{\rho}{G}}v^4 \vec{\nabla} \left( s + \ln\rho\right)\,.
\label{Fourierlaw s}
\ee
This equation is supplemented by the continuity~\eqref{final continuity}, momentum~\eqref{final momentum} and Poisson~\eqref{final poisson} 
equations.
 
To simplify the analysis, at this point we approximate mass and entropy densities as nearly uniform, thereby neglecting their
gradients:~$\vec{\nabla}s, \vec{\nabla}\rho \simeq 0$. In other words, we treat~$\rho$ and~$s$ as average quantities. It follows 
from~\eqref{Fourierlaw s} that the heat flux can also be neglected,~$\vec{q} \simeq 0$. Hence~\eqref{entropy eqn}
simplifies to
\be
\frac{\partial s}{\partial t}  =  \frac{\dot{{\cal E}}}{mv^2}\,.
\label{1st law}
\ee
Not surprisingly, the entropy of DM particles increases as they are heated by baryons.

Assuming that the initial DM entropy (at virialization) is negligible compared to its final value (at equilibrium),~\eqref{1st law} can be
schematically integrated over a relaxation time to give
\be
\frac{\dot{{\cal E}}}{mv^2}\, t_{\rm relax} \sim  1 \,.
\ee
This expresses the condition for equilibrium. Substituting~\eqref{heat rate final} and~\eqref{t relax fix}, 
we obtain 
\be
\frac{\Sigma_{\rm DM}^3}{v^2} \sim  \frac{a_0 \rho_{\rm b}}{G^2} \,,
\label{sigma v2}
\ee
where we have used~\eqref{R v rho0} and~\eqref{sigma DM} to estimate the DM surface density as~$\Sigma_{\rm DM} \sim \sqrt{\frac{\rho v^2}{G}}$. 

Meanwhile, we know that the central baryonic surface density of an exponential disk of scale-length~$L$ is ~$\Sigma_{\rm b}(0) = \frac{M_{\rm b}}{2 \pi L^2}$. Assuming an approximate linear relation $L_{\rm z} \approx L/8$ between disk scale-length and scale-height, we can approximate the mean baryon density by~$\rho_{\rm b} \sim \tfrac{M_{\rm b}}{L^3} \sim \tfrac{\Sigma_{\rm b}^{3/2}(0)}{\sqrt{M_{\rm b}}}$. Substituting into~\eqref{sigma v2}, we obtain
\be
\frac{\Sigma_{\rm DM}^3}{v^2} \sim \frac{\left(\frac{a_0 \Sigma_{\rm b}(0)}{G}\right)^{3/2}}{\sqrt{a_0GM_{\rm b}}}\,.
\label{sigma v2 final}
\ee
Hence, taking the BTFR~$v^2 \sim {\sqrt{a_0GM_{\rm b}}}$ as a given, we get
\be
\Sigma_{\rm DM} \propto \sqrt{\frac{a_0 \Sigma_{\rm b}(0)}{G}} \,.
\label{Sigma LSB}
\ee
This is the desired CSDR, valid for DM-dominated (LSB) galaxies. Because the analysis relied on the {\it average} density, it is
not surprising that the result matches the DM-dominated CSDR. On the other hand, we have already seen
within the equilibrium treatment that such a relation holds for HSB galaxies. 

It will be important to quantify the numerical coefficient in~\eqref{sigma v2 final}, as well as its scatter. This will require numerical simulations of galaxy
formation within our scenario, which is beyond the scope of the present analysis. It is nevertheless encouraging that the correct parametric
dependence of the scaling relations derives from a back-of-the-envelope analysis. 

\section{Cosmological Implications and Constraints}
\label{cosmo}

In this Section we consider a few astrophysical and cosmological implications of our model. We will be able to derive
very general results, using only the form of the heating rate~\eqref{heat rate final}, without specifying an explicit
microphysical model. The analysis does rely, however, on the assumption that the physics underlying our DM-baryon interactions still apply in the various
environments studied below, such as in the early universe. For instance, if heat transport is due to collective excitations of a DM medium ({\it e.g.}, fluid or solid), 
our working assumption is that this DM condensed state is a valid description in these environments. 

For comparison with the constraints below, we will set~$C = \tfrac{1}{10}$ for concreteness and assume~$a_0 = 10^{-8}~{\rm cm}/{\rm s}^2$. Our heating rate~\eqref{heat rate final} 
then becomes
\be
\frac{\dot{\cal E}}{m}  = 10^{-9}\, \frac{\rho_{\rm b}}{\rho} v~\frac{{\rm cm}}{{\rm s}^2} \,.
\label{heat predict}
\ee
Thus the predicted heating rate is determined simply by the DM-to-baryon fraction and velocity dispersion in the relevant environments.

\subsection{Early universe}

DM-baryon interactions can affect the evolution in the early universe. In the case of interest where baryons heat up DM, the dominant constraint comes spectral
distortions of the CMB taking place in the redshift range~$10^4 \;\lsim\; z \;\lsim\; 10^6$~\cite{Ali-Haimoud:2015pwa}. In the standard cosmological model, baryons are kept in thermal equilibrium with photons by Compton scattering until~$z \simeq 200$. This process effectively cools photons, causing small spectral distortions. This cooling will be enhanced if baryons shed part of their thermal
energy to DM, resulting in larger and potentially observable spectral distortions.  

This effect was studied in detail in the case of light DM ($m \ll m_{\rm b}$) scattering elastically with baryons and/or photons~\cite{Ali-Haimoud:2015pwa}. It is straightforward to
translate their result to a constraint on the energy exchange rate~$\dot{\cal E}$. Consider the energy exchange rate per baryon,~$\tfrac{\dot{\cal E}n}{n_{\rm b}}$, relative to the thermal energy~$\sim m_{\rm b}v_{\rm b}^2$ per baryon, where~$n_{\rm b}$ and~$v_{\rm b}$ are respectively the baryon number density\footnote{For the purpose of this simple estimate, we ignore the distinction between nuclei and free electrons.} and velocity dispersion. Let us compare this to the Hubble rate by defining
\be
\epsilon \equiv \frac{\dot{\cal E}n/n_{\rm b}}{Hm_{\rm b}v_{\rm b}^2} = \frac{C}{6} \frac{a_0 v}{Hv_{\rm b}^2}\,,
\ee
where the last step follows from~\eqref{heat rate final}. 

The effect on spectral distortions will be negligible if~$\epsilon \ll 1$ in the redshift range~$10^4 \;\lsim\; z \;\lsim\; 10^6$. 
It is easy to check that~$\epsilon$ increases in time in this range, hence the constraint is most stringent at~$z \simeq 10^4$.
Since baryons are in thermal equilibrium with radiation, we have~$v_{\rm b}^2 = T_\gamma/m_{\rm b}$, with~$T_\gamma$
denoting the CMB temperature. Substituting~$T_\gamma \simeq 2 {\rm eV}$ and~$H \simeq 10^{-27}~{\rm eV}$ at~$z \simeq 10^4$,
together with our fiducial values~$C = \tfrac{1}{10}$ and~$a_0 = 10^{-8}~{\rm cm}/{\rm s}^2$, we obtain
\be
\left.\epsilon\right\vert_{z = 10^4} \simeq 10\,\frac{v}{c}\,.
\ee
Since our DM particles are assumed non-relativistic at that time,~$v\ll c$, the resulting spectral distortions are indeed negligible.

\subsection{Merging clusters}

Merging galaxy clusters constrain the DM self-interaction cross section per unit mass~\cite{Markevitch:2003at,Randall:2007ph,Harvey:2015hha,Wittman:2017gxn},
\be
\frac{\sigma}{m} \;\lsim\; \frac{{\rm cm}^2}{{\rm g}}\,.
\label{bullet sigma}
\ee
The precise numerical value of the coefficient depends on the assumptions, but is~${\cal O}(1)$ or less~\cite{Harvey:2015hha,Wittman:2017gxn}. This can be translated to
a constraint on the heating rate of DM per unit mass,~$\frac{\dot{\cal E}}{m}  \simeq \rho \frac{\sigma}{m} v^3$, where we have used a characteristic energy exchanged
per collision of~$m v^2$ for DM-DM scattering. Substituting the characteristic density~$\rho \simeq 10^{-24} {\rm g}/{\rm cm}^3$ and
velocity~$v \simeq 10^3~{\rm km}/{\rm s}$ for merging clusters, the bound~\eqref{bullet sigma} translates to 
\be
\frac{\dot{\cal E}}{m}  \;\lsim\; \frac{{\rm cm}^2}{{\rm s}^3}\,.
\label{mergingclusterbound}
\ee

Although~\eqref{bullet sigma} was derived assuming DM self-interactions, the end result applies equally well to our heating rate obtained from  
DM-baryon scattering. Substituting into~\eqref{heat predict} the DM-baryon ratio~$\rho \sim 10\, \rho_{\rm b}$ in clusters and relative velocity
$v \simeq 10^3~{\rm km}/{\rm s}$, we obtain
\be
\frac{\dot{\cal E}}{m}\bigg\vert_{{\rm clusters}}  \simeq 10^{-2} \;\frac{{\rm cm}^2}{{\rm s}^3}\,.
\ee
This comfortably satisfies~\eqref{mergingclusterbound}. On the flip side, a couple order of magnitude improvement in the observational bound~\eqref{bullet sigma} would probe
our predicted heating rate, thereby highlighting the power of merging clusters for detecting DM-baryon interactions.

\subsection{Cosmic Dawn and the EDGES anomaly}
\label{edges sec}

The recent measurement of the 21-cm absorption spectrum from the Cosmic Dawn epoch by the EDGES collaboration revealed an excess signal~\cite{Bowman:2018yin}.
If real, the excess could indicate that the hydrogen gas at~$z \simeq 17$ was cooler than predicted by the standard~$\Lambda$CDM
model. A possible explanation is that interactions between DM and baryons acted to cool the neutral gas prior to the Cosmic
Dawn~\cite{Barkana:2018lgd}. 

For instance, sub-GeV DM particles scattering elastically with baryons with velocity-dependent cross section,
\be
\sigma_{\rm int}(v) = \sigma_1 \left(\frac{v}{1~{\rm km}/{\rm s}}\right)^{-4}\,,
\label{sigma v4}
\ee
%s
would explain the signal if
\be
\sigma_1 \;\gsim\; 10^{-20}~{\rm cm}^2\,.
\label{sigma1 bound}
\ee
The strong velocity dependence of~\eqref{sigma v4} is necessary to evade cosmological and astrophysical bounds~\cite{Barkana:2018lgd,Tashiro:2014tsa,Munoz:2015bca}. 
Detailed model-building analyses, however, show that it is difficult to construct explicit particle physics models
that are compatible with other constraints~\cite{Munoz:2018pzp,Berlin:2018sjs,Barkana:2018cct,Kovetz:2018zan}. 

Equations~\eqref{sigma v4} and~\eqref{sigma1 bound} can be translated to a heating rate per unit mass using~$\frac{\dot{\cal E}}{m} \simeq n_{\rm b} \sigma_{\rm int}(v) v^3$.
Substituting the cosmological baryon number density~$n_{\rm b} = 2 \times 10^{-7}(1+z)^3~{\rm cm}^{-3}$ evaluated at~$z\simeq 17$,
together with the characteristic velocity~$v = 1~{\rm km}/{\rm s}$, the bound~\eqref{sigma1 bound} translates to
\be
\frac{\dot{\cal E}}{m} \;\gsim\; 10^{-8}~\frac{{\rm cm}^2}{{\rm s}^3}\,.
\label{edges bound}
\ee
This is how large the heating rate ought to be to explain the EDGES excess.
In our case, substituting into~\eqref{heat predict} the cosmological ratio~$\rho \simeq 6\,\rho_{\rm b}$, together with
$v = 1~{\rm km}/{\rm s}$, our predicted heating rate is
\be
\frac{\dot{\cal E}}{m}\bigg\vert_{z = 17} \simeq 2\times 10^{-5}~\frac{{\rm cm}^2}{{\rm s}^3}\,.
\ee
Thus our heating mechanism can explain the EDGES excess.

\section{Conclusions}
\label{conclusions}

Among the small-scale challenges of~$\Lambda$CDM~\cite{Bullock:2017}, the conspiracy between DM and baryon distributions in disk galaxies, embodied in the MDAR, 
is arguably one of the most tantalizing. The MDAR is a unique relation between the total gravitational field and the Newtonian acceleration generated by baryons alone at
every radius in disk galaxies. In particular, both the tightness of the BTFR and the diversity of galaxy rotation curves that it implies~\cite{Ghari:2018jkc} remain challenging within the~$\Lambda$CDM framework, where this conspiracy must arise through feedback processes. While semi-empirical arguments based on abundance matching can reproduce the general shape of the MDAR, its normalization, and especially its very small scatter, remain challenging~\cite{Li:2018tdo}. Relatedly, it has recently been pointed out that stellar feedback is related to  a characteristic acceleration of order~$a_0$. While promising, this is not sufficient yet to explain the details of the diversity of rotation curves encoded in the tightness of the MDAR, which should be related to the subtleties of the core-cusp transformation process.  On the numerical front, much progress has been made in obtaining the MDAR from hydrodynamical simulations of galaxy formation, as reviewed in the Introduction, though challenges -- related to the extreme tightness of the BTFR and diversity of rotation curves -- still remain. 

Given these challenges, it is worthwhile to entertain the alternative possibility that the baryon-DM conspiracy embodied by the MDAR is due to new, 
non-gravitational interactions between the two sectors. Traditionally, work in this direction has focused on postulating a new long-range {\it force} acting on
baryons, thereby effectively modifying gravity. This force could be either fundamental or, as in superfluid DM, emergent from the DM medium. 

The idea pursued in this paper, building on our earlier work~\cite{Famaey:2017xou}, is that the MDAR is the result of {\it direct} (non-gravitational) interactions between DM and baryons, instead of an effective modification of gravity or feedback processes. The main difference with our earlier work is to consider that this interaction {\it heats} the DM-fluid. The approach followed has been
completely ``bottom-up". Using a hydrodynamical description of DM, our goal has been to identify which such DM-baryon interactions are necessary to reproduce the MDAR. 

In this framework, the microphysics of DM is encoded in three physical quantities: the DM equation of state,~$P = P(\rho,v)$; the
relaxation time,~$t_{\rm relax}$, which enters in the heat conductivity; and the energy exchange rate~$\dot{\cal E}$, which is
determined by DM-baryon interactions. A key result of this work is that the MDAR is obtained if the following conditions are satisfied:

\begin{enumerate}

\item The equation of state is approximately that of an ideal gas,~$P = \rho v^2$. This will generically be realized in the dilute limit, where
the average inter-particle separation is large compared to the mean free path. 

\item The relaxation time is set by the Jeans time,~$t_{\rm relax} \sim \tfrac{1}{\sqrt{G\rho}}$. This can be achieved naturally, for instance, if DM is in a Knudsen regime~\cite{Famaey:2017xou}.

\item The heating rate satisfies the master relation~$\tfrac{\dot{{\cal E}}}{m} \sim C a_0 v \tfrac{\rho_{\rm b}}{\rho}$. This is the most important relation as it informs us about the necessary DM-baryon particle interactions. 

\end{enumerate}  

To be clear, we do not claim that these are unique nor necessary, but they are {\it sufficient} to obtain the MDAR. Remarkably, with these assumptions the set of hydrodynamical equations, together with Poisson's equation, enjoy an anisotropic scaling symmetry, which offers yet another guide for model building. Moreover, in DM-dominated regions this scaling symmetry is enhanced to a one-parameter family of scalings, implying the  scaling relation~\eqref{eq:LSBscaling}, which fully captures the low-acceleration limit of the MDAR.

In this paper, we built on and further developed the original scenario of~\cite{Famaey:2017xou} in several crucial ways. Most importantly, as stated above, instead of baryon-DM interactions cooling the DM medium, we focused exclusively on the case where the DM fluid is {\it heated} by baryons. This is indeed {\it a priori} more desirable from the point of view of galaxy formation, since DM heating can transform cusps into cores in the central regions of galaxy halos. It also avoids the concern of forming flattened halos or dark disks. A second key difference pertains to the form of DM-baryon interactions. Whereas our original analysis~\cite{Famaey:2017xou} focused exclusively on short-range particle-particle collisions between DM and baryons, in the present analysis we
remained general about the form of such interactions. This opens up a wider range of possibilities for particle physics model-building.

We then showed how, assuming a cored pseudo-isothermal profile, the above hydrodynamical ingredients give rise at equilibrium to suitable parameters reproducing the MDAR.
Specifically, in the flat part of the rotation curve the asymptotic rotational velocity matches the parametric dependence of the BTFR, up to a logarithm in~$r$. Meanwhile, in the
central region of HSB galaxies, where baryons dominate, the DM profile reproduces the CSDR with the behaviour of the 'simple' interpolating function of MOND. Finally, by studying the time-dependent approach to equilibrium, we derived a constraint on a combination of the DM velocity dispersion and surface density, which matches the combination of BTFR and CSDR. Therefore, if one takes the BTFR as a given (per the equilibrium analysis), this constraint yields
the CSDR naturally.

Remarkably, the form of the heating rate makes definite, model-independent predictions for various cosmological and astrophysical observables.
The only assumption of course is that the underlying DM-baryon effective theory responsible for the heating rate is still valid in these different
environments. Assuming this is the case, we argued that our model satisfies various observational constraints, and, intriguingly, offers a
possible explanation to the EDGES excess. Of course, there will be many more phenomenological loops to go through once we have an
explicit particle physics realization, but it is reassuring that our heating rate so far appears to be observationally viable.

Our framework offers a number of avenues for further development. Three particularly important directions are:

\begin{itemize}

\item {\bf Including the dynamics of baryons:} In our framework we focused our attention on the dynamics of the DM sector, treating baryons as an external source. This is a reasonable approximation provided that the typical energy lost by a baryon is not significant enough to affect its dynamics over the time scales of interest. Using the expression~\eqref{heat rate final} for our heating rate~$\dot{\mathcal{E}}$, one can estimate the energy lost by a baryon per unit length to be~$\tfrac{{\rm d} E_{\rm b}}{{\rm d} \ell} \gtrsim \tfrac{C  m_{\rm b} a_0 v}{V_{\rm b}}$. Even keeping in mind that~$C \sim \mathcal{O}(10^{-1})$, this quantity could become large enough in some LSBs, and a more accurate treatment would require including the dynamics of baryons.  

\item {\bf Numerical simulations of galaxy formation:} Our scenario is ripe for a fully dynamical study of galaxy formation. Because our equations are cast
in simple hydrodynamical terms, it should be straightforward to modify existing hydrodynamical codes to include our heating rate. For this purpose, the
formulation in terms of entropy density presented in Sec.~\ref{approach to equi} may be most convenient. Such numerical studies would inform us, among other things, on the stability of the equilibrium solution, in particular whether the outskirts of galaxy disks are not too severely perturbed by interactions with DM. It would allow us to check whether the equilibrium configuration is reached dynamically on the predicted time scale. Furthermore, such an analysis would also allows us to quantify the expected scatter for the BTFR, in particular for the characteristic scale~$R_0$ appearing in the logarithm. 

\item {\bf Building a particle physics model:} In this paper we have adopted a purely bottom-up approach based on an effective hydrodynamical description of the DM sector. It would be very interesting to deduce what type of constraints the heating rate~\eqref{heat rate final} poses on the underlying microscopic interactions between baryons and DM. One promising way of ensuring that our scenario is compatible with small-scale ({\it e.g.}, solar system) constraints would be to consider interactions that involve collective excitations emerging at scales of~$\mathcal{O}$(pc). We leave the exploration of this interesting possibility for future work. 

\end{itemize}

\vspace{.4cm}
\noindent
{\bf Acknowledgements:} We thank Lasha Berezhiani and Scott Dodelson for helpful discussions. B.F. acknowledges support from the Agence Nationale de la Recherche (ANR) project ANR-18-CE31-0006, and from the European Research Council (ERC) under the European Union's Horizon 2020 research and innovation programme (grant agreement No. 834148). J.K. is supported in part by the US Department of Energy (HEP) Award DE-SC0013528, NASA ATP grant 80NSSC18K0694, and a W.~M.~Keck Foundation Science and Engineering Grant. R.P. is supported in part by the National Science Foundation under Grant No. PHY-1915611.

\renewcommand{\em}{}
\bibliographystyle{utphys}
\addcontentsline{toc}{section}{References}
\bibliography{heating_new_v6}

\end{document}